\documentclass{article}

\usepackage{arxiv}

\usepackage[utf8]{inputenc} 
\usepackage[T1]{fontenc}    
\usepackage{hyperref}       
\usepackage{url}            
\usepackage{booktabs}       
\usepackage{amsfonts}       
\usepackage{microtype}      
\usepackage{lipsum}
\usepackage{graphicx}
\graphicspath{ {./images/} }

\usepackage{subcaption} 
\usepackage{xspace}
\usepackage{amsmath}
\usepackage{siunitx}
\usepackage{multirow}
\usepackage{adjustbox}

\title{deepEOSnet: Capturing the dependency on thermodynamic state in property prediction tasks}

\author{
  Jan Pav\v{s}ek \textsuperscript{1}, \quad Alexander Mitsos \textsuperscript{1,2,3}, \quad Manuel Dahmen \textsuperscript{3}, \quad Tai Xuan Tan \textsuperscript{1}, \quad Jan G. Rittig \textsuperscript{1*} \vspace*{2mm}\\
	\textsuperscript{1}{Process Systems Engineering (AVT.SVT), RWTH Aachen University} \\
	\textsuperscript{2}{JARA Center for Simulation and Data Science (CSD)}\\
	\textsuperscript{3}{Institute of Climate and Energy Systems ICE-1: Energy Systems Engineering, Forschungszentrum Jülich GmbH}\\
	\textsuperscript{*}{Corresponding author, \texttt{jan.rittig@rwth-aachen.de}}\\
}

\begin{document}

\maketitle
\begin{abstract}

We propose a machine learning (ML) architecture to better capture the dependency of thermodynamic properties on the independent states.
When predicting state-dependent thermodynamic properties, ML models need to account for both molecular structure and the thermodynamic state, described by independent variables, typically temperature, pressure, and composition.
Modern molecular ML models typically include state information by adding it to molecular fingerprint vectors or by embedding explicit (semi-empirical) thermodynamic relations.
Here, we propose to rather split the information processing on the molecular structure and the dependency on states  into two separate network channels: a graph neural network and a multilayer perceptron, whose output is combined by a dot product.
We refer to our approach as DeepEOSNet, as this idea is based on the DeepONet architecture [Lu et al. (2021), Nat. Mach. Intell.]: instead of operators, we learn state dependencies, with the possibility to predict equation of states (EOS).
We investigate the predictive performance of DeepEOSNet by means of three case studies, which include the prediction of vapor pressure as a function of temperature, and mixture molar volume as a function of composition, temperature, and pressure. 
Our results show superior performance of DeepEOSNet for predicting vapor pressure and comparable performance for predicting mixture molar volume compared to state-of-research graph-based thermodynamic prediction models from our earlier works.
In fact, we see large potential of DeepEOSNet in cases where data is sparse in the state domain and the output function is structurally similar across different molecules.
The concept of DeepEOSNet can easily be transferred to other ML architectures in molecular context, and thus provides a viable option for property prediction.

\end{abstract}

\section{Introduction}

\noindent Molecular machine learning (ML) has advanced thermodynamic property prediction, promising to accelerate the identification of molecules with desired properties and thereby more efficient chemical processes~\cite{rittig2025molecular}.
Various ML methods, such as graph neural networks (GNN)~\cite{santana2024puffin, rittig2024thermodynamics_consistent, lansford2023physics,hoffmann2025grappa}, transformers~\cite{winter2023spt_nrtl,winter2025understanding,winter2022smile} and matrix completion methods~\cite{damay2021mcmTdep} have shown high predictive accuracy for various thermodynamic properties of both pure components and mixtures, such as normal boiling points~\cite{Rittig_GNNBook.2022}, solubility~\cite{Vermeire.2021}, and activity coefficients~\cite{damay2021mcmTdep, rittig2024thermodynamics_consistent, winter2023spt_nrtl, medina2022graph}.
The predictive capabilities of these methods have been further improved by utilizing thermodynamic knowledge within model training and/or as part of the model architecture~\cite{rittig2023gibbs_informed,rittig2024thermodynamics_consistent,hoffmann2025grappa,santana2024puffin,lansford2023physics}, 
thereby ensuring thermodynamic consistency while preserving ML model flexibility, cf. overview in~\cite{rittig2025molecular}.
Therefore, molecular ML can provide property predictions with high accuracy respecting thermodynamic principles and is highly promising for chemical engineering applications.

For state-dependent thermodynamic properties, it is critical to also account for information on the independent state variables, typically how they vary with temperature, pressure, and composition.
As such, molecular ML models have been extended in various cases, including the prediction of vapor-liquid equilibria as function of composition and pressure/temperature~\cite{sun2025vapor,medina2025graph}, vapor pressures as function of temperature~\cite{santana2024puffin,lansford2023physics,hoffmann2025grappa,lin2024Antoine,winter2025understanding}, activity coefficients as function of composition~\cite{rittig2023gibbs_informed, rittig2024thermodynamics_consistent}, temperature~\cite{rittig2023graph,winter2022smile,damay2021mcmTdep,medina2023gibbs}, and both~\cite{winter2023spt_nrtl,specht2024hanna}, density as function of temperature~\cite{felton2024ml_saft,winter2025understanding}, Helmholtz free energy as function of density and temperature~\cite{chaparro2024developmentMLeos,chaparro2023mie_fluid}, and Gibbs free energy as function of temperature~\cite{hammad2025gnnGibbsstruct}. 
Two different approaches to capture the state dependency are frequently used in molecular ML:
\begin{itemize}
	\item[] \emph{Concatenation}: 
	The state variables are concatenated to the molecular fingerprint vectors, learned by the ML models, and are then considered as an additional input to a multilayer perceptron (MLP) for property prediction, e.g., in~\cite{rittig2023graph,brozos2025predicting,aouichaoui2023s}.
	\item[] \emph{Embedding of semi-empirical equations}:
	Semi-empirical models that provide consecutive equations of states are incorporated into the ML architecture, in the form of hybrid models, e.g., in~\cite{medina2023gibbs, winter2023spt_nrtl,hoffmann2025grappa}.
\end{itemize}
A challenge of simply concatenating molecular fingerprint vectors and state information, as in the \emph{concatenation} approach, is that the MLP has to simultaneously learn the molecular structure and state dependency of the property of interest purely from the provided data, and thus in some sense distinguish between the two effects.
Also physics-informed ML models that use fundamental thermodynamics have to learn the state dependency purely from data:
While fundamental thermodynamics provides algebraic/differential relationships (e.g., Gibbs-Helmholtz equation and Maxwell relations) between properties and stability criteria that all state functions must satisfy in equilibrium, they do not specify the state dependency of properties for specific molecules/mixtures explicitly.
The determination of state dependencies -- whether $(\partial{C_p}/\partial{P})|_T$ is positive or negative, or how fugacity coefficients vary with temperature -- requires molecular structure-specific information, and is typically based on statistical mechanics, molecular simulations, or experimental measurements.
Notably, in physics-informed ML models using fundamental thermodynamics, learning the state dependencies is not only needed for making state dependent predictions but also for enforcing the fundamental relationships accurately.
As for the \emph{embedding of semi-empirical equations} approach, semi-empirical models typically parametrize the state dependency in few molecule- or mixture-specific parameters, thereby significantly reducing the number of parameters that need to be learned in ML model training (in comparison to an MLP).
This potentially reduces data demands for training, but also relies on simplifying assumptions, such as using ideal gas law for the vapor phase in the Antoine equation~\cite{pfennig2013thermodynamik}, hence comes with the limitations of semi-empirical models.

We propose an alternative molecular ML approach to simultaneously learn the molecule- and the state dependency of thermodynamic properties, without introducing modeling limitations and thus preserving ML model flexibility. 
To this end, we do not rely on a semi-empirical thermodynamic model as in the \emph{embedding of semi-empirical equations} approach; we rather propose a modification to the \emph{concatenation} approach, inspired by DeepONets (deep operator networks)~\cite{lu2021deeponet}. 
We refer to our architecture as DeepEOSNet, as we learn state functions corresponding to equations of state (EOS) instead of operators, and therefore replace ``O'' by ``EOS''.

Lending from the DeepONet structure, in DeepEOSNet we separate the learning of the molecular structure and state dependencies into two different channels.
The molecular channel encodes structure-specific information into the molecular fingerprint vector that is relevant for the property of interest; it can thus be based on any molecular ML model, such as transformers and GNNs.
The state channel captures the general functional form of the influence of the independent state variables on the property.
The outputs of the two channels are then combined by a dot product to give the final property prediction.
In this work, we apply our approach to learning individual properties, i.e., vapor pressure and total molar volume of mixtures.
We stress that the DeepEOSNet approach is also directly transferable to learning fundamental properties, e.g., Gibbs free energy, as introduced by Rittig et al. \cite{rittig2024thermodynamics_consistent}. 

\section{Fundamentals} \label{sec:fundamentals}

\noindent We first describe the concepts of DeepONets for operator learning and GNNs for molecular property prediction, on which we then build our proposed DeepEOSNet architecture.

\subsection{DeepONet for operator learning}

\begin{figure}[h]  
	\centering
	\includegraphics[trim = 72 92 82 32, clip ,width=\textwidth]{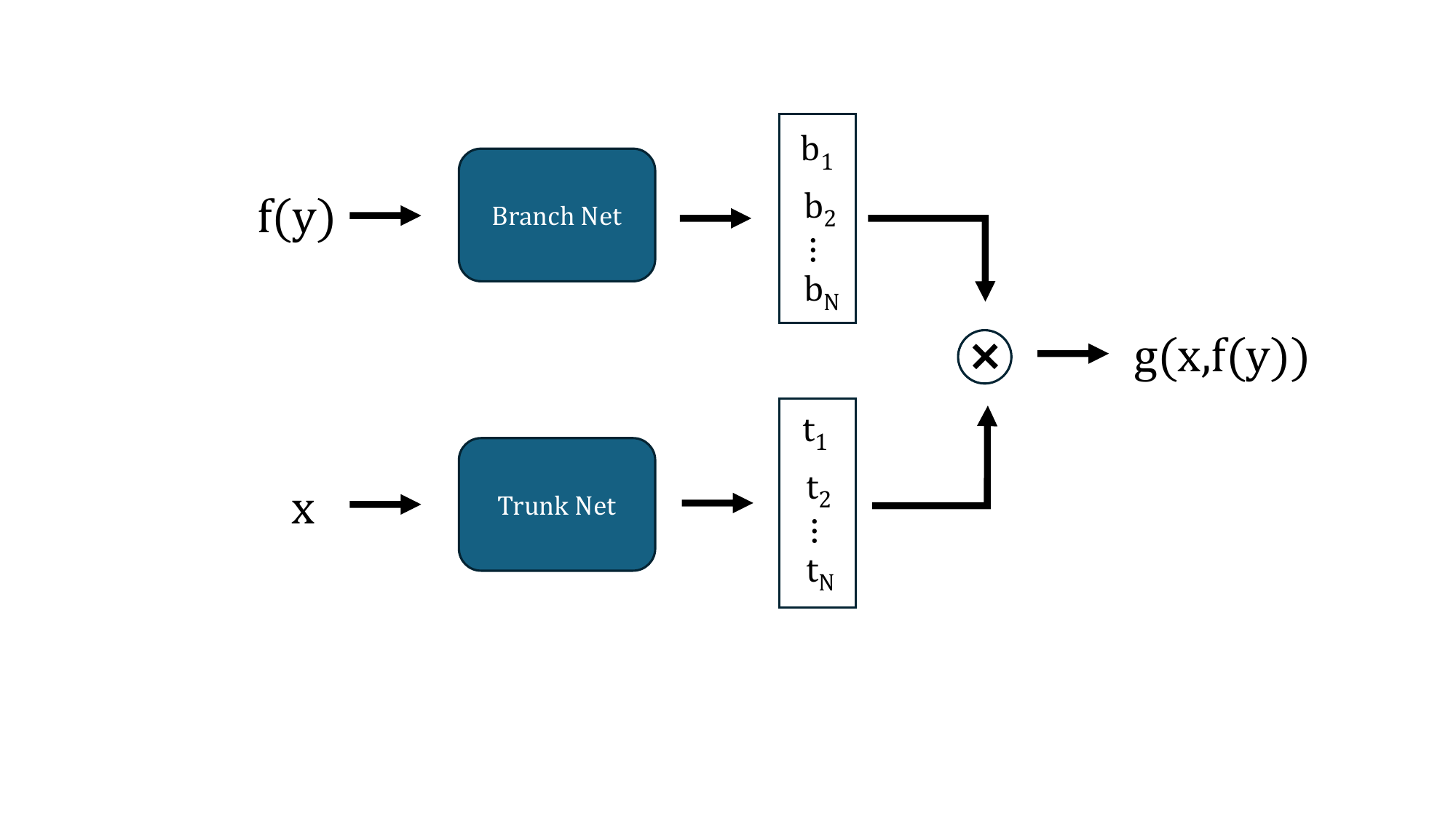}  
	\caption{Schematic illustration of the DeepONet architecture}
	\label{fig:arch_DeepONet}
\end{figure}

\noindent DeepONets, introduced by Lu et al.~\cite{lu2021deeponet}, are a deep neural network architecture designed to learn operators, i.e., a mapping from one function ($f(\cdot)$) to another ($g(\cdot)$), which they apply to ordinary differential equations (ODEs) and partial differential equations (PDEs).
The DeepONet architecture is illustrated in Figure~\ref{fig:arch_DeepONet}: Two separate neural networks, referred to as branch and trunk net, are employed for splitting information processing.
The branch net processes the information of the input function, whereas the trunk net processes information on the domain of the output function. 
The results of these networks are then combined by a dot product, such that the output can be described by

\begin{equation} \label{eq:deepONet}
	g(x,f(y)) = \sum_{i=1}^{N_{hd}} t_{i}(x)b_{i}(f(y)),
\end{equation}
where $g(\cdot)$ denotes the output function $N_{hd}$ is the hidden dimension of the branch and trunk net, $b_{i}$ is the output of the i$-th$ node of the branch net, $t_{i}$ is the output of the i$-th$ node of the trunk net, $f(\cdot)$ is the input function, $y$ is the independent variable of the input function, and $x$ is the independent variable of the output function. 
In the example of PDEs, the branch net processes the PDE with specific initial/boundary conditions, whereas the trunk net processes the evaluation points.
As such, DeepONets can learn to solve the PDE for different input conditions, rather than at a single condition, hence they learn solution operators.

Practically, such DeepONets approximate the function mapping by predicting a target value $g(x,f(y))$ that is dependent on the two input features $f(y)$ and $x$. 
At the same time, correlation between $f(y)$ and $x$ might be present, in case $x$ and $y$ are correlating or data sampling is unequally distributed across the domain.
Separating the information processing into branch and trunk net appears to mitigate the effects potential correlations in the data points of $f(\cdot)$ and $x$ might have on training. 
The separate networks ensure that the influence each of the inputs $f(y)$ and $x$ have on the target $g(x,f(y))$ is separated from the other input.
At the same time this architecture is more versatile during inference, as the evaluation of $g(x,f(y))$ at new domain points $x$ only requires passing through the trunk net in case $f(y)$ is constant, and vice versa.
Notably, separate processing of inputs has also been explored in other domains, such as image/video processing using SplitNets~\cite{dong2022splitnets}.

\subsection{Graph neural networks for molecular property prediction}

\begin{figure}[b!]  
	\centering
	\includegraphics[trim = 12 222 92 142, clip ,width=\textwidth]{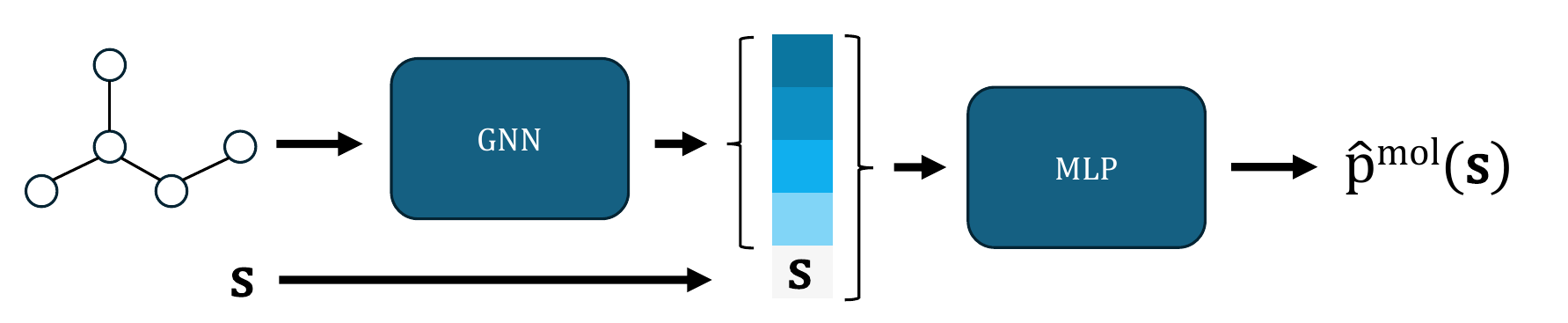}  
	\caption{Architecture of a graph neural network for predicting state-dependent thermodynamic properties}
	\label{fig:arch_GNN4thermo}
\end{figure}

\noindent GNNs learn properties directly from a graph representation of molecules, as illustrated in Figure~\ref{fig:arch_GNN4thermo}.
For a detailed overview of GNN architectures in the molecular context, we refer the interested reader to~\cite{schweidtmann2020graph,rittig2023graph, Reiser.2022, Gilmer.2017}.
The general working principle of GNN is as follows:
Graph convolutional layers first encode structural information from the molecular graph, which results in a continuous vector representation of the molecule, referred to as molecular fingerprint vector. 
This molecular fingerprint vector is then passed through an MLP which provides a prediction for the target property $\hat{\text{p}}$.

In case the prediction target additionally depends on at least one independent thermodynamic state variable -- typically temperature, pressure, and/or composition --, these additional state variables $\mathbf{s}$ need to be accounted for in addition to the molecular structure. 
Most commonly, the state variables $\mathbf{s}$ are concatenated with the molecular fingerprint vector~\cite{rittig2023graph,aouichaoui2023s}, hence, they serve as additional input to the MLP.
The MLP then simultaneously learns the molecular structure and state dependencies.

\section{DeepEOSNet for thermodynamic properties} \label{sec:den}

\begin{figure}[h]  
	\centering
	\includegraphics[trim = 0 92 82 32, clip ,width=\textwidth]{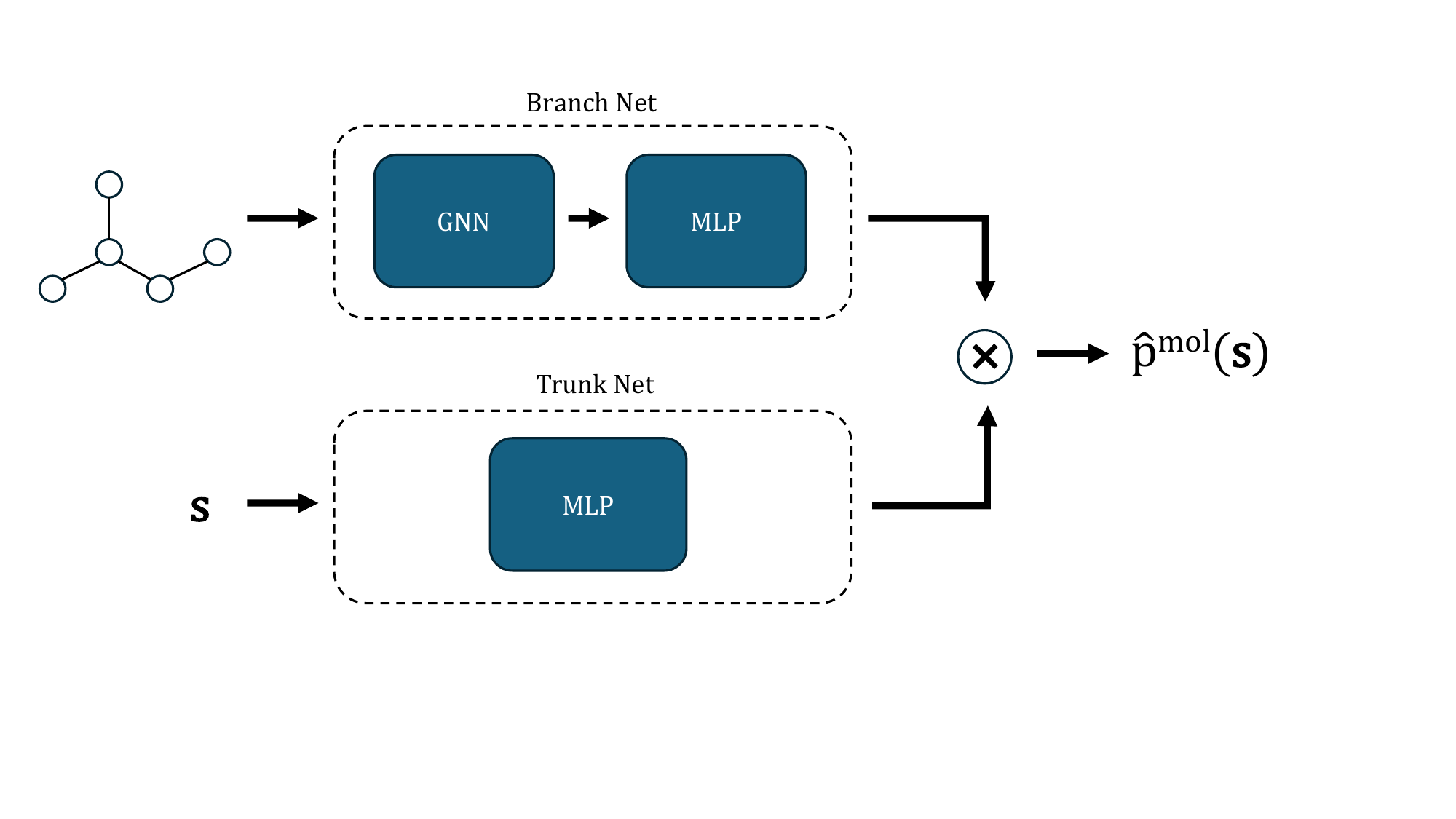}
	\caption{Architecture of DeepEOSNet}
	\label{fig:arch_DEN}
\end{figure}

\noindent We propose DeepEOSNet, an ML architecture for state-dependent thermodynamic property prediction.
DeepEOSNet splits the information processing of the molecular structure input and the state variables into two separate network channels, as shown in Figure~\ref{fig:arch_DEN}. 
The branch net is a GNN, that extracts structural information from the molecular graph, resulting in a molecular fingerprint vector. 
The trunk net is an MLP, which takes the state variable $\mathbf{s}$ as input and outputs a continuous vector with the same dimension as the molecular fingerprint vector.
The two output vectors are then combined by a dot product to yield the final property prediction $\hat{\text{p}}^{\text{mol}}(\mathbf{s})$.

By separating the information processing of the molecular structure and state variables, a stronger inductive bias is introduced to learn the influences on the molecular property separately:
The state channel (trunk net) does not have the molecular structure as an input, and thus learns a general functional form of the property as function of states.
Vice versa, the molecule channel (branch net) does not have direct access to the state variables, and therefore extracts molecular structure-specific information.
We stress that the molecule channel can still learn any state-dependent information.
In fact, during model training, the molecule channel can learn from state-dependent differences between the predicted and actual values (i.e., the experimental or simulated property values). 
This allows the molecule channel to capture the molecule-specific state dependency within the general form of state dependency induced by the state channel.

Interestingly, employing a dot product based calculation of the final output (see Equation~\ref{eq:den}) is conceptually similar to the commonly employed representation using summation over simple basis functions  used in thermodynamics to capture the dependency on independent states.
A well-known example of this is the NASA polynomial representation~\cite{mcbride2002nasa} (see Equation~\ref{eq:NASA_pol}) of thermodynamic properties, which is commonly used to capture temperature dependency:
\begin{align}
	\hat{\text{p}}^{\text{mol}}_{\text{DeepEOSNet}}(\mathbf{s}) &= \sum_{i=1}^{N_{hd}} t_{i}(\mathbf{s})*b_{i}(G^{\text{mol}}) \label{eq:den}\\
	\hat{\text{p}}^{\text{mol}}_{\text{NASA}}(T) &= a_{1}T^{-2} + a_{2}T^{-1} + a_{3} + a_{4}T + a_{5}T^{2} + a_{6}T^{3} + a_{7}T^{4} \label{eq:NASA_pol}
\end{align}
where $G^{\text{mol}}$ is a molecular graph, $T$ is the temperature and $a_{i}$ are exemplary molecule parameters for a seventh-order NASA polynomial.
Specifically, the temperature polynomials correspond to $t_{i}(\mathbf{s})$ and the molecule-specific parameters to $b_{i}(G^{\text{mol}})$. 
Therefore, the output layer of the DeepEOSNet architecture is essentially a learnable and more general form of the NASA polynomial, as $t_{i}(\mathbf{s})$ is not constrained to polynomials but rather the MLP which is a universal function approximator.

Notably, the same functional behavior, as given by the structure of the DeepEOSNet, can also be learned by the molecular ML architectures that simply concatenate the state variable to the molecular fingerprint vector (see Figure~\ref{fig:arch_GNN4thermo} and approach (1) in the introduction).
Yet, constraining the network architecture in DeepEOSNet is expected to facilitate training, as it reduces the search space of functions that can be learned for each molecule specifically to the general functional form induced by the state channel (trunk net)~\cite{wilson2020bayesian}.
We hypothesize that this is particularly helpful for properties for which the general functional form of the state dependency can be easily approximated by the state channel, which is likely if state dependencies are similar for the different molecules of interest.

We further note that Sun et al.~\cite{sun2023deepgraphonet} have proposed DeepGraphONet which modifies DeepONet by employing a GNN as branch net, and is thus similar to our architecture.
However, in their application, namely the prediction of system behavior of power grids and traffic flow, the input to the branch net $f(\cdot)$ (see Figure~\ref{fig:arch_DeepONet}) is a graph which itself is a function of the input to the trunk net $\mathbf{x}$. 
That is, the graph structure itself depends on the output domain, in this case the time.
In contrast, in DeepEOSNet, the input to the branch net, which is the molecular graph, does not depend on the state feature vector, cf. Figure~\ref{fig:arch_DEN}. 
Therefore, the input to the branch net in DeepEOSNet is not a function.
We thus rather learn state \emph{functions} instead of operators, over the discrete space of molecules/mixtures.
Further, as the focus of our approach is the application of separate information processing to state dependencies in molecular prediction tasks in general, we refer to our approach as DeepEOSNet.

\section{Property prediction case studies}

\noindent For testing the DeepEOSNet approach, we consider two target properties of high practical relevance for chemical engineering applications, namely vapor pressure and total molar volume of binary mixtures.
Note that traditionally the total molar volume of mixtures is calculated by summing up the pure component molar volumes weighted with their mole fractions and adding the excess volume of mixing (since the ideal mixing volume is zero).
However, here we directly predict the total molar volume of the mixture.
The reason for this is that we want to avoid the introduction of modeling assumptions that potentially limit predictive performance.
Predicting first the behavior of ideal mixtures and then the excess volume might introduce an inductive bias for the model towards simplified ideal mixture behavior.
To avoid this, we directly train on the total molar volume of the mixture in an end-to-end fashion. 
In the following, we use the terms total molar volume of the mixture and mixture molar volume interchangeably.

\subsection{Dataset}

\noindent We extract property data from the NIST ThermoData Engine~\cite{nist_tde}. 
The \textbf{vapor pressure} dataset contains 879 molecules with vapor pressures at varying temperatures in the range from 78\,K to 875\,K, amounting to 71,730 data points in total. 
Notably, the number of data points with different temperatures is unevenly distributed among the molecules, ranging from a few data points for some molecules to thousands for others.
The \textbf{molar volume} dataset contains 265 molecules combined into 1,444 unique binary mixtures with varying temperatures in the range from 201\,K to 505\,K, pressures in the range from 78\,kPa to 291\,MPa, and compositions between 0 and 1. 
The total number of molar volume data points is 88,365.
Again, the distribution of data points is rather uneven for the different mixtures, in particular with respect to temperature and pressure variation.

As both datasets exhibit largely varying numerical values for the property data points, the logarithm of pressure and molar volume, respectively, is used as prediction target. 
The temperature and pressure input are linearly normalized to range 0-1. 
We thus report the error metrics in Section~\ref{sec:res_n_dis} on the logarithmic scale.

\subsection{Prediction scenarios \& baselines} \label{sec:case_studies}

\noindent Using the vapor pressure and mixture volume datasets, we create three thermodynamic property prediction case studies (see Table~\ref{tab:over_casestudies}).

\begin{table}[h]
	\centering
	\caption{Overview case studies}
	\begin{tabular}{c|c|c|c|c}
		
		No. & Abbreviation & $\mathbf{s}$ & $\hat{\text{p}}^{\text{mol}}(\mathbf{s})$ & Train set\\
		\hline
		\multirow{2}{*}{1} & \multirow{2}{*}{$\text{p}^{sat}_{extra}\text{(T)}$} & \multirow{2}{*}{$T$} & \multirow{2}{*}{$p^{sat}(T)$} & lowest $80\%$ temperature \\ 
		& & & & points per molecule \\
		\hline
		\multirow{2}{*}{2} & \multirow{2}{*}{$\text{p}^{sat}_{SingleT}\text{(T)}$} & \multirow{2}{*}{$T$} & \multirow{2}{*}{$p^{sat}(T)$} & one temperature \\ 
		& & & & point per molecule  \\
		\hline
		\multirow{2}{*}{3} & \multirow{2}{*}{$\text{V}^{mix}_{SingleX}\text{(T,p,x)}$} & \multirow{2}{*}{$[T, p, x]^{T}$} & \multirow{2}{*}{$V^{mix}(T, p, x)$} & one composition point \\
		& & & & per mixture, $T$ and $p$ \\
	\end{tabular}
	\label{tab:over_casestudies}
\end{table}

In the first two prediction tasks, the temperature-dependent vapor pressure $p^{sat}(T)$ is selected as target property.
Specifically, we consider two scenarios:
\begin{itemize}
	\item[]$\text{p}^{sat}_{extra}\text{(T)}$: We select the $80\%$ lowest temperature points for each molecule as training set and use the highest $20\%$ as test set. For molecules where there are less than 5 temperature points in the dataset, all points are sorted to the training set.
	In doing so, we test the extrapolation capabilities in the temperature domain, i.e., to a temperature range not used in training at all, which serves as proxy for learning temperature dependency in general. 
	\item[]$\text{p}^{sat}_{SingleT}\text{(T)}$: We randomly select a single temperature point per molecule for training, while all remaining temperature points are sorted to the test set. 
	This allows us to test the capabilities of models to distinguish between the molecular structure and state dependencies, as the variance of the two input dependencies overlaps entirely in this scenario.
	We note that this scenario does presumably not represent a practical scenario for experimental datasets and induces an unrealistic difficulty but is insightful to test the methodological model characteristics.  
\end{itemize}

For the third prediction task, we consider the total molar volume of binary mixtures as a function of temperature, pressure, and composition. 
This scenario adds difficulty to the prediction in two ways: First, considering binary mixtures requires considering two molecular graphs as model input instead of a single one. Secondly, accounting for three state dependencies results in a three dimensional state variable vector $\mathbf{s}$ instead of a one-dimensional one as in the vapor pressure scenarios.
This third prediction task is denoted as follows:

\begin{itemize}
	\item[]$\text{V}^{mix}_{SingleX}\text{(T,p,x)}$: We perform a train-test split such that each binary mixture is only available at a single composition (but at varying temperatures and pressures) during training, whereas all data points corresponding to the remaining compositions of this binary mixtures are used for testing.
	Notably, this scenario represents a more realistic case compared to the $\text{p}^{sat}_{SingleT}\text{(T)}$, as density measurements for mixtures might be more sparse in the composition domain. 
\end{itemize}

\noindent In all scenarios, we employ a GNN using concatenation of the state variables to the molecular fingerprint vectors as a baseline, denoted as GNN$_\text{concat}$.
For the two vapor pressure prediction scenarios, we also consider as a baseline a GNN that embeds the Antoine equation, which is a hybrid model that predicts the parameters of the Antoine equation in the output head, similar to~\cite{lansford2023physics,santana2024puffin,hoffmann_jirasek2025grappa,lin2024Antoine}.
We refer to this as GNN$_\text{Antoine}$.

\subsection{Implementation \& Hyperparameter}

\noindent We implement DeepEOSNet and the GNN baselines in Python using PyTorch Geometric~\cite{FeyLenssen2019pytorchgeom}, based on our graph ML framework for molecular and mixture property prediction 
\href{https://git.rwth-aachen.de/avt-svt/public/gmolprop/}{\emph{GMoLprop}}.
All GNN model parts share the same basic architecture. 
The GNN baseline that embeds the Antoine equation includes adaptations following the approach by Hoffmann et al.~\cite{hoffmann2025grappa}: A tanh activation function is applied in the last layer before entering the Antoine output head, and training follows a two step procedure, that is, first training with mean squared error (MSE) loss and then switching to the Huber loss~\cite{hoffmann2025grappa}.

We optimize model hyperparameters in a two-step grid search approach. 
First, the hyperparameters concerning the model architecture are optimized for all three models. 
Then, with fixed optimized architectural hyperparameters, the training parameters are determined. 
The optimized hyperparameters are then used to generate the results for all three case studies.
In the first step, the optimized parameters are the dimension of the molecular fingerprint~$\in~\{64, 128\}$, the number of MLP layers~$\in~\{1, 2, 3\}$, the activation function in the branch net~$\in~\{\text{LeakyReLU, ReLU}\}$ and in the trunk net~$\in~\{\text{tanh, sigmoid}\}$, and the latent feature dimension $N_{hd} \in \{20,30,40\}$. 
During the first step, the learning rate is fixed to 0.001, the batch size to 64, and the drop out rate to 0.
For the DeepEOSNet, the hyperparameter optimization results in a fingerprint dimension of 128, 3 MLP layers in the branch net, LeakyReLU as activation in the GNN/branch net and tanh in the trunk net, and a latent feature dimension of 40.
The GNN with state variable concatenations exhibits a fingerprint dimension of 128, 3 MLP layers, and the LeakyReLU activation function.
For the GNN$_\text{Antoine}$, the fingerprint dimension is 128, the number of MLP layers is 3, and LeakyReLU is chosen as the activation function.
In the second step, the initial learning rate~$\in~\{0.01, 0.001, 0.0001\}$, the batch size~$\in~\{32, 64, 128\}$ and the dropout rate~$\in~\{0.1, 0.05, 0\}$ are optimized.
This yields the same training procedures for all three models, with a learning rate of 0.001, a batch size of 64, and a dropout rate of 0, which correspond to the hyperparameters chosen in the first step.
Furthermore, we train all models for 300 epochs with early stopping, and repeat training 10 times with different seeds.

\section{Results and Discussion} \label{sec:res_n_dis}

\noindent We first present and discuss the results from the vapor pressure scenarios (Section~\ref{sec:vap_pres_res}) and then consider the more difficult mixture molar volume scenario (Section~\ref{sec:mol_v_res}). 
For all numeric performance metrics and training curves reported in the subsequent sections, the averages and standard deviations across the 10 different seeds are reported.
Parity plots and individual molecule examples are shown for the best run, in terms of performance on the test set, of each model across the 10 different seeds.

\subsection{Vapor pressure prediction}
\label{sec:vap_pres_res}

\noindent Table~\ref{tab:perform_met_vap_pres} provides an overview of the prediction test performance of DeepEOSNet and the baseline GNN models, GNN$_\text{concat}$ and GNN$_\text{Antoine}$, for the two vapor pressure case studies.  
In general, both the GNN$_\text{concat}$ and the DeepEOSNet achieve high prediction accuracy across both vapor pressure prediction scenarios with $R^{2}$ values above $0.9$. 
The GNN$_\text{Antoine}$ exhibits higher RMSE and MAE values, and an average $R^{2}$ below 0.
Thus, in both case studies, both the GNN$_\text{concat}$ and the DeepEOSNet significantly outperform the GNN$_\text{Antoine}$.

\begin{table}[htpb]
	\centering
	\caption{Performance metrics, root mean squared error (RMSE), mean absolute error (MAE), and coefficient of determination ($\text{R}^{2}$) evaluated on the logarithmic scale, of different prediction models on test set for the two vapor pressure prediction scenarios.}
	\label{tab:perform_met_vap_pres}
	\begin{adjustbox}{center}
		\begin{tabular}{c c|c c c|c c c|c c c}
			
			\multirow{2}{*}{Scenario} & & \multicolumn{3}{c|}{$\text{GNN}_{\text{concat}}$} & \multicolumn{3}{c|}{DeepEOSNet} & \multicolumn{3}{c}{$\text{GNN}_{\text{Antoine}}$}  \\
			& & RMSE & MAE & $\text{R}^{2}$ & RMSE & MAE & $\text{R}^{2}$ & RMSE & MAE & $\text{R}^{2}$ \\
			\hline
			
			\multirow{2}{*}{$\text{p}^{sat}_{extra}\text{(T)}$}& avg & 0.12 & 0.045 & 0.98 & 0.13 & 0.060 & 0.98 & 0.75 & 0.41 & $<$ 0 \\
			& $\sigma$ & 0.008 & 0.006 & 0.002 & 0.011 & 0.010 & 0.003 & 0.73 & 0.34 & 1.8 \\
			\hline
			\multirow{2}{*}{$\text{p}^{sat}_{SingleT}\text{(T)}$}& avg & 0.34 & 0.21 & 0.95 & 0.31 & 0.20 & 0.96 & 8.9 & 0.64 & $<$ 0 \\
			& $\sigma$ & 0.032 & 0.026 & 0.010 & 0.049 & 0.047 & 0.014 & 16 & 0.34 & 420 \\
			\hline
			
		\end{tabular}
	\end{adjustbox}
\end{table}

Notably, the performance metrics reported here for the GNN$_\text{Antoine}$ are larger than those reported by Hoffmann et al.~\cite{hoffmann2025grappa}, with an MAE of 0.41 compared to 0.2. 
However, Hoffmann et al. use a different train-test split, sorting unseen molecules to the test set, whereas we test for extrapolation capabilities in the temperature domain. 
As extrapolation to unseen molecules is an inherently different task than extrapolation to unseen temperature regions, the metrics can not be directly compared.

\begin{figure}[thpb]
	\centering
	
	\begin{subfigure}[t]{0.48\textwidth}
		\centering
		\includegraphics[width=\linewidth]{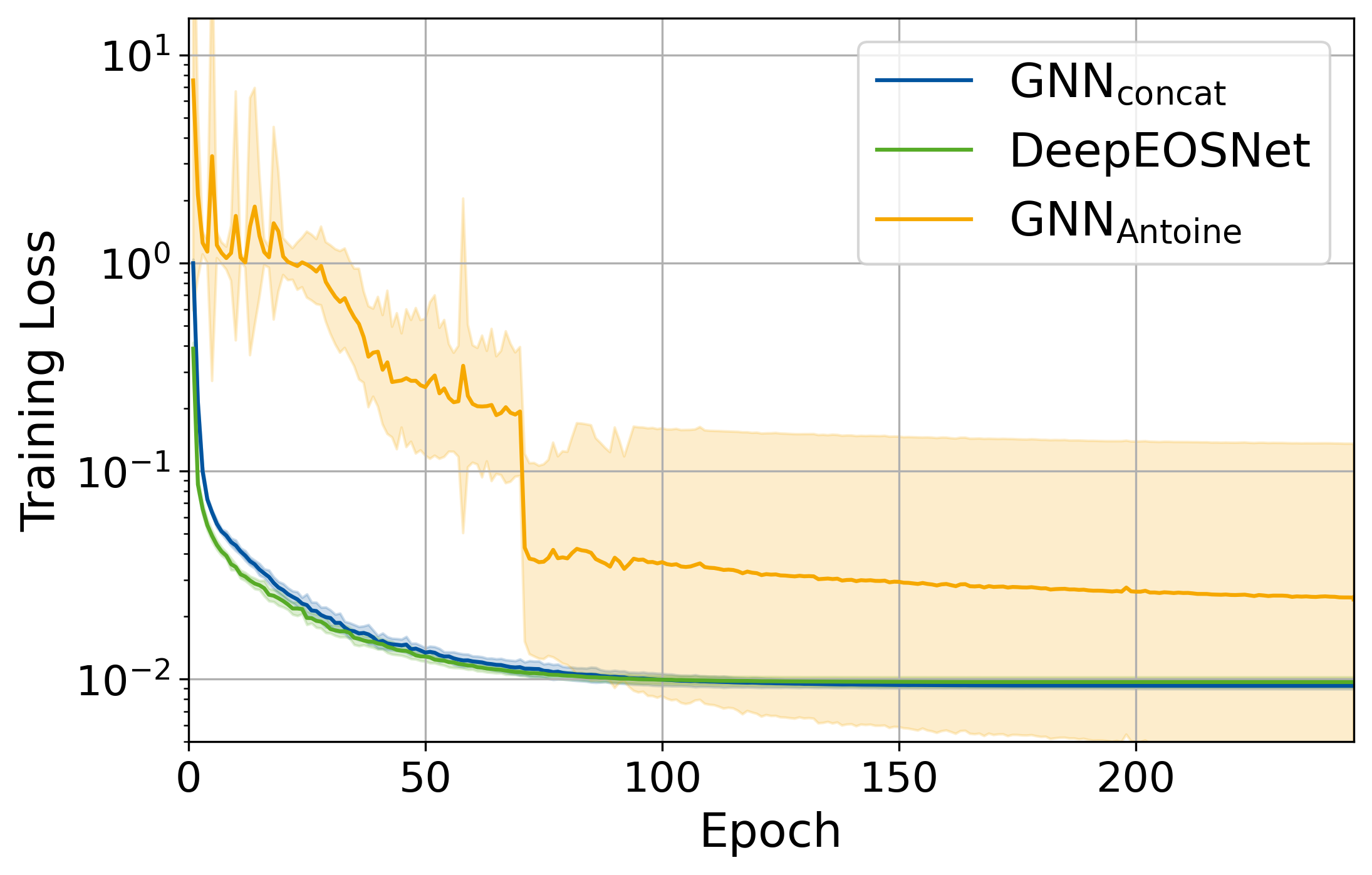}
		
		\includegraphics[width=\linewidth]{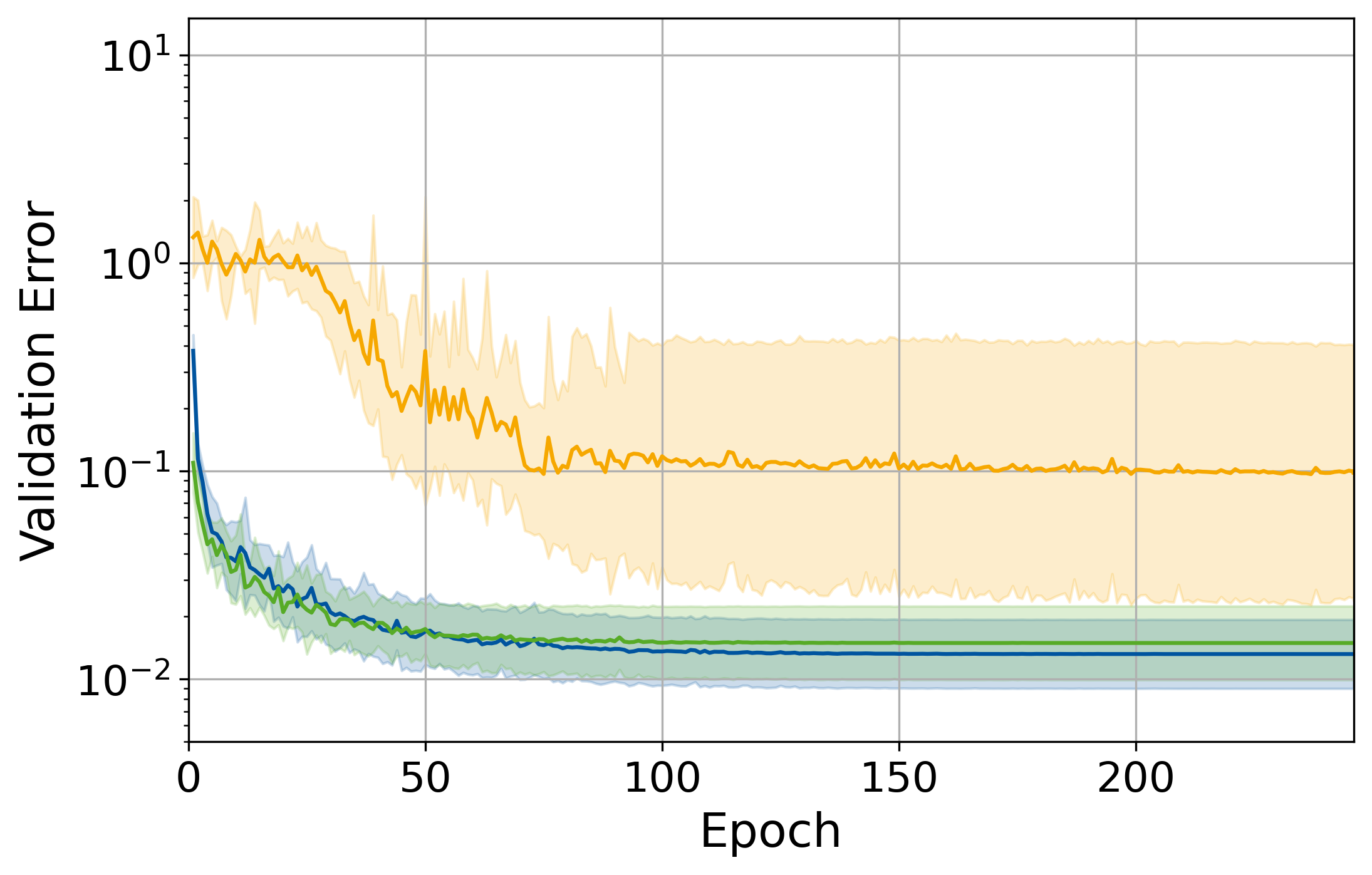}
		\caption{$\text{p}^{sat}_{extra}\text{(T)}$}
		\label{fig:train_curve_tempExt}
		
	\end{subfigure}
	\hfill
	\begin{subfigure}[t]{0.48\textwidth}
		\centering
		\includegraphics[width=\linewidth]{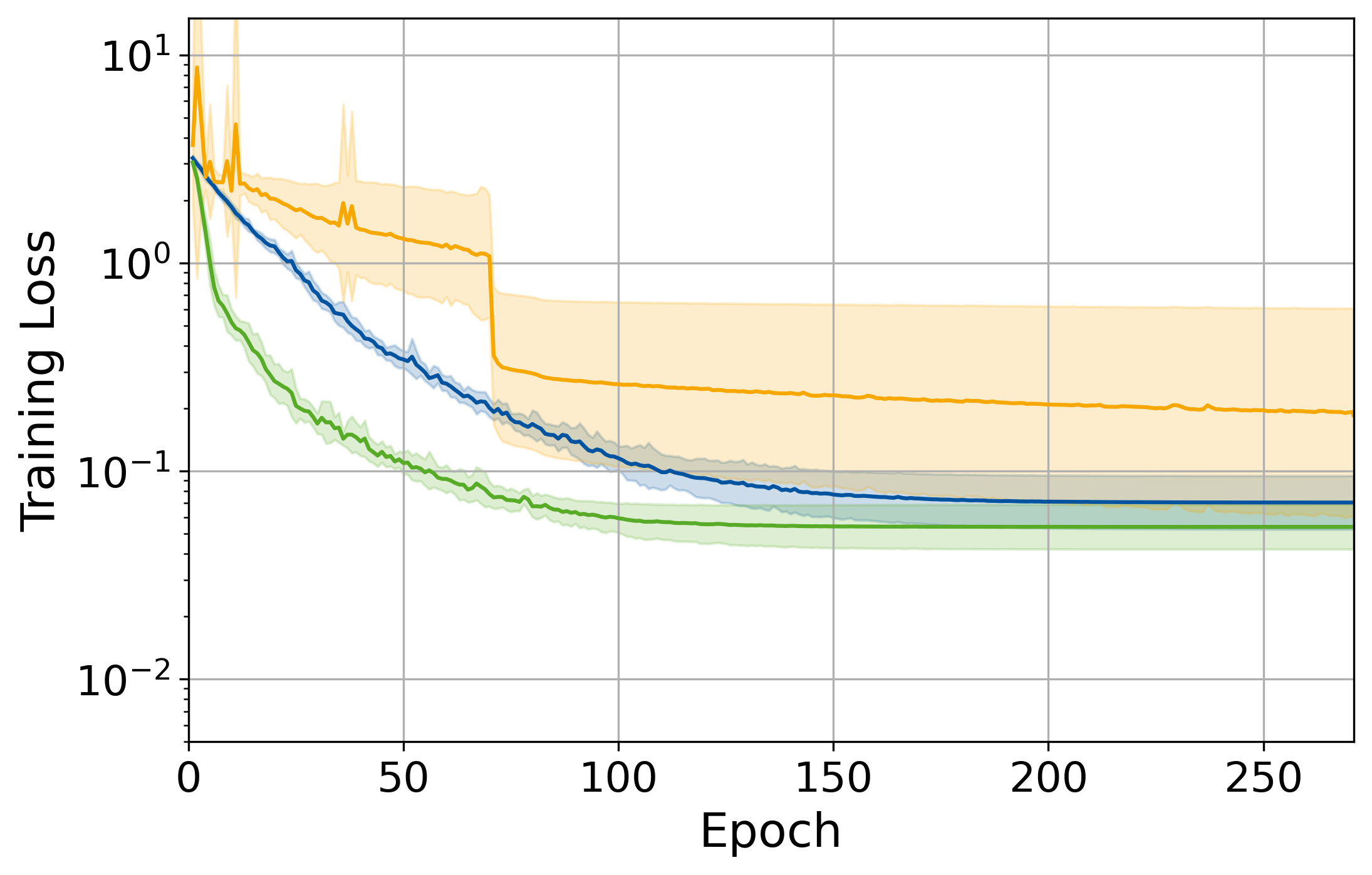}
		
		\includegraphics[width=\linewidth]{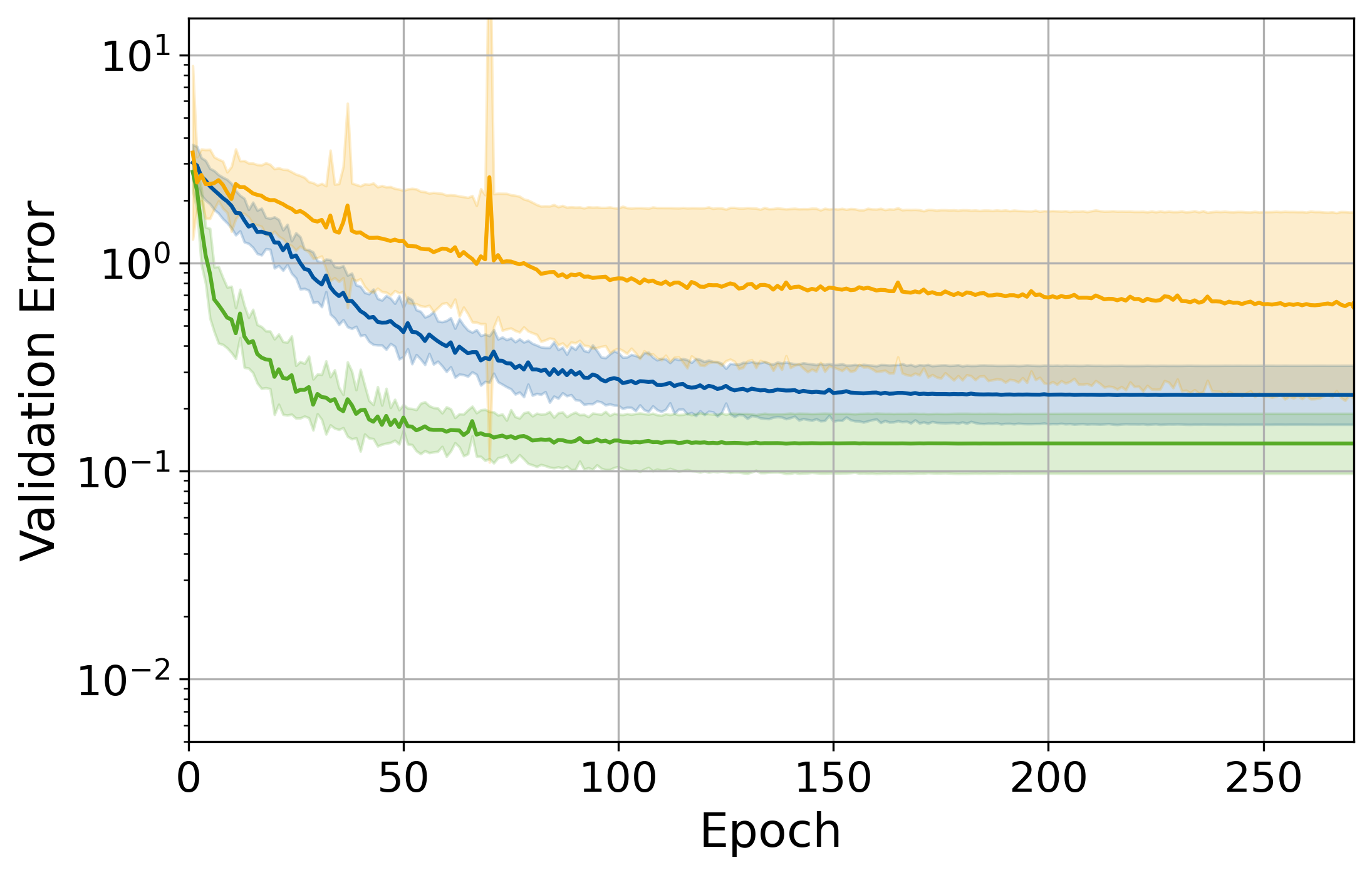}
		\caption{$\text{p}^{sat}_{SingleT}\text{(T)}$}
		\label{fig:train_curve_sTpM}
	\end{subfigure}
	
	\caption{Training curves}
	\label{fig:train_curve}
\end{figure}

Here, we generally find a high variance in prediction performance of the GNN$_\text{Antoine}$.
In fact, despite the low average $R^{2}$ values (see Table~\ref{tab:perform_met_vap_pres}), it achieves $R^{2}$ values above 0.9 for some of the ten runs, which is reflected in the high standard deviation and indicates training instabilities.
The GNN$_\text{concat}$ and DeepEOSNet show reasonably low standard deviation in the performance metrics.
In Figure~\ref{fig:train_curve}, we show the training and validation loss for both vapor pressure scenarios, where the line represents the average and the shaded area the standard deviation between the 10 runs.
We find that both the GNN$_\text{concat}$ and the DeepEOSNet have significantly faster convergence and, indeed, more stable training across different seeds than the GNN$_\text{Antoine}$, so the ease of training is an advantage of both the GNN$_\text{concat}$ and DeepEOSNet.

To further compare the model capabilities, we analyze the two vapor pressure prediction scenarios individually in the following.

\pagebreak

\subsubsection{\textbf{$\text{p}^{sat}_{extra}\text{(T)}$ scenario}} \vspace{2mm}

\begin{figure}[hbtp]
	\centering
	
	\begin{subfigure}{0.33\textwidth}
		\centering
		\includegraphics[width=\linewidth]{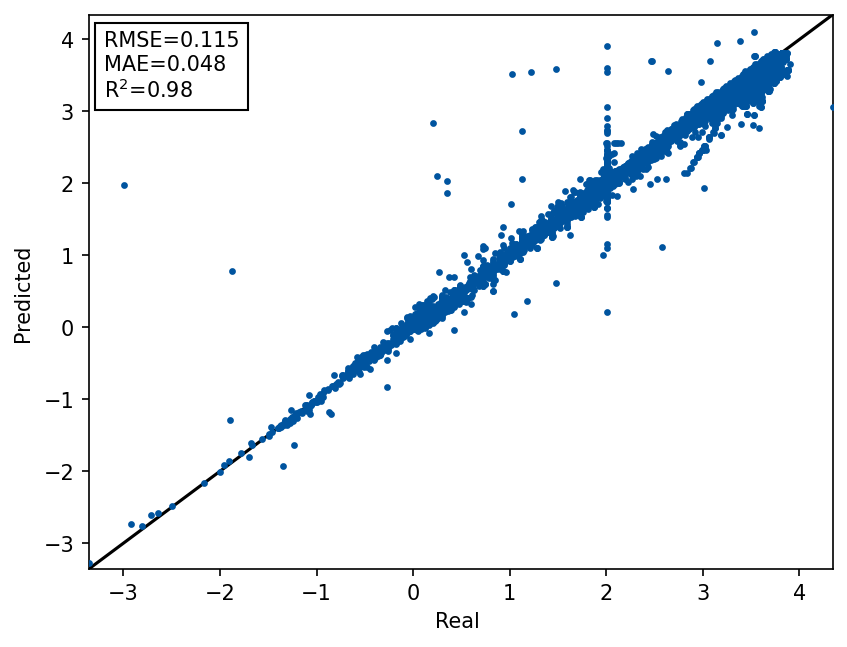}
		\caption{DeepEOSNet}
	\end{subfigure}
	\begin{subfigure}{0.33\textwidth}
		\centering
		\includegraphics[width=\linewidth]{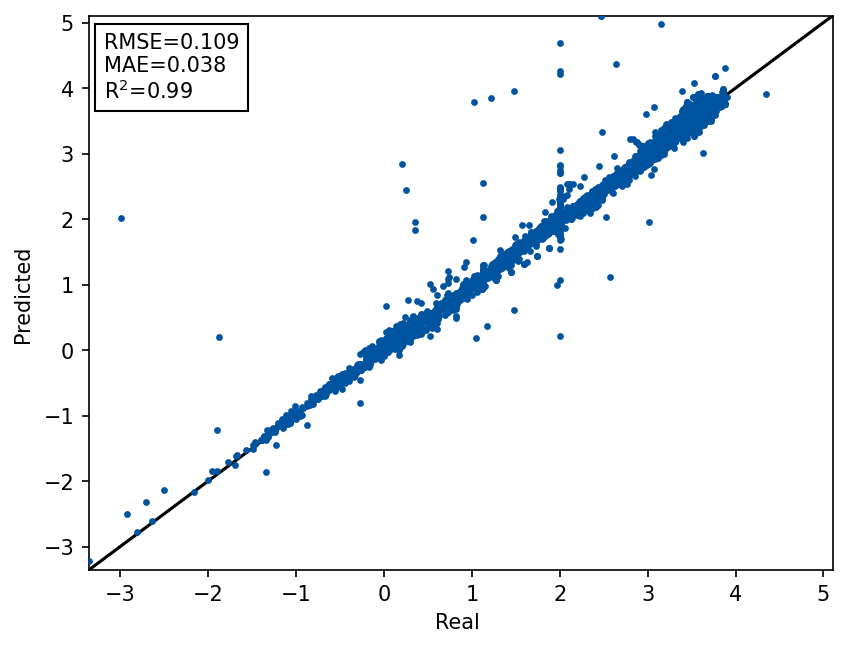}
		\caption{$\text{GNN}_{\text{concat}}$}
	\end{subfigure}
	\begin{subfigure}{0.33\textwidth}
		\centering
		\includegraphics[width=\linewidth]{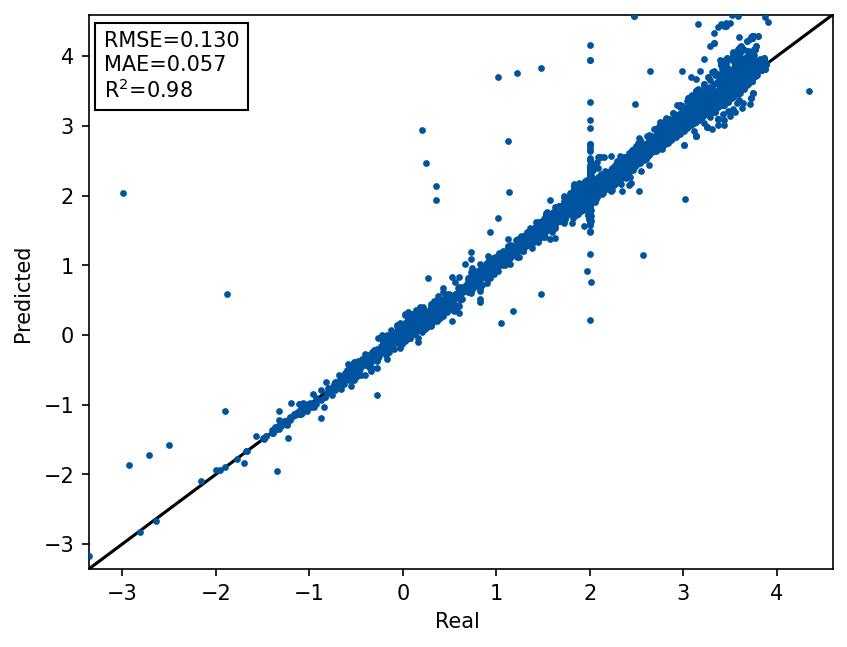}
		\caption{$\text{GNN}_{\text{Antoine}}$}
		\label{fig:par_plot_tempext_GNN_ant}
	\end{subfigure}	
	
	\caption{Parity plots of test set $\text{p}^{sat}_{extra}\text{(T)}$}
	\label{fig:parity_p_text}
\end{figure}

\noindent In Figure~\ref{fig:parity_p_text}, we show parity plots for the $\text{p}^{sat}_{extra}\text{(T)}$ scenario, i.e., excluding all vapor pressure data points in a higher temperature range (cf. Section~\ref{sec:case_studies}).
We stress that the results in the parity plots corresponds to the best out of ten different training runs for each model (based on test error), allowing to compare the models for the case of stable training of the GNN$_\text{Antoine}$ model.

Both the GNN$_\text{concat}$ as well as the DeepEOSNet exhibit a tight distribution along the diagonal, indicated by the similar RMSE of 0.115 for the DeepEOSNet and 0.109 for the GNN$_\text{concat}$.
For the GNN$_\text{Antoine}$ the best performing run achieves comparative performance to the other two models, with an $\text{R}^{2}$ of 0.98, which is significantly above its average (see Table~\ref{tab:perform_met_vap_pres}). 
Only in the high vapor pressure region, the distribution is slightly wider than in the other two models with the general tendency to overestimate (see Figure~\ref{fig:par_plot_tempext_GNN_ant}).
In the high vapor pressure region, also the DeepEOSNet slightly deviates from the diagonal with a tendency to underestimate.
The GNN$_\text{concat}$ remains in its trend closer to the diagonal in the high vapor pressure region.
Furthermore, for all three models, numerous outliers are visible; however, these do not indicate general patterns in mispredictions. 
As they seem to be present in equal amounts for all models, we rather hypothesize that these outliers can be partly explained by noise and erroneous points in the experimental data used for training. 
Overall, the results of the $\text{p}^{sat}_{extra}\text{(T)}$ case suggest that the DeepEOSNet as the more restrictive architecture poses neither a benefit nor a disadvantage for predictive performance.
In fact, all models show the capability to predict vapor pressures at temperatures higher than those used in training with reasonable accuracy.

\subsubsection{\textbf{$\text{p}^{sat}_{SingleT}\text{(T)}$ scenario}} \vspace{2mm}

\begin{figure}[h]
	\centering
	\begin{subfigure}[t]{0.33\textwidth}
		\centering
		\includegraphics[width=\linewidth]{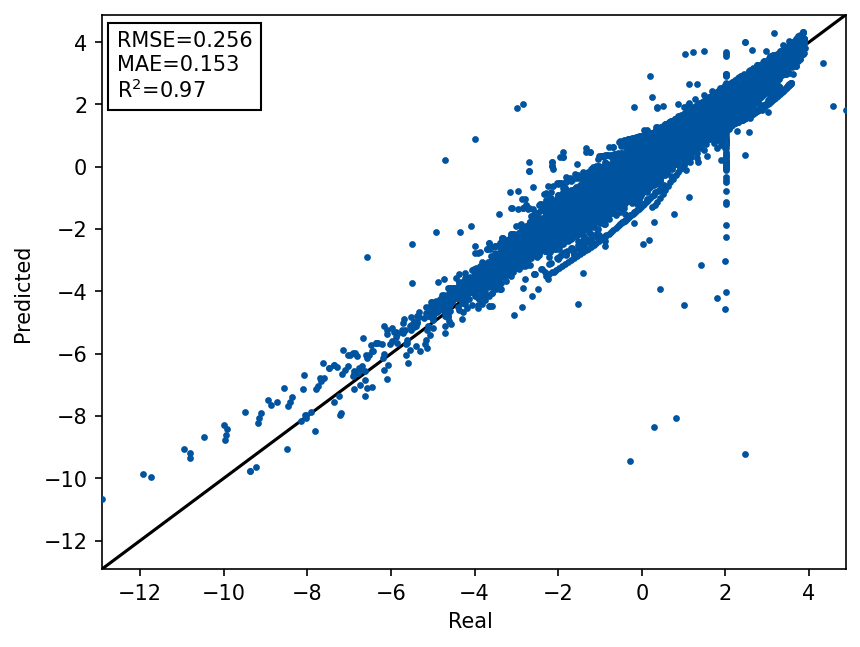}
		\caption{DeepEOSNet}
	\end{subfigure}
	\begin{subfigure}[t]{0.33\textwidth}
		\centering
		\includegraphics[width=\linewidth]{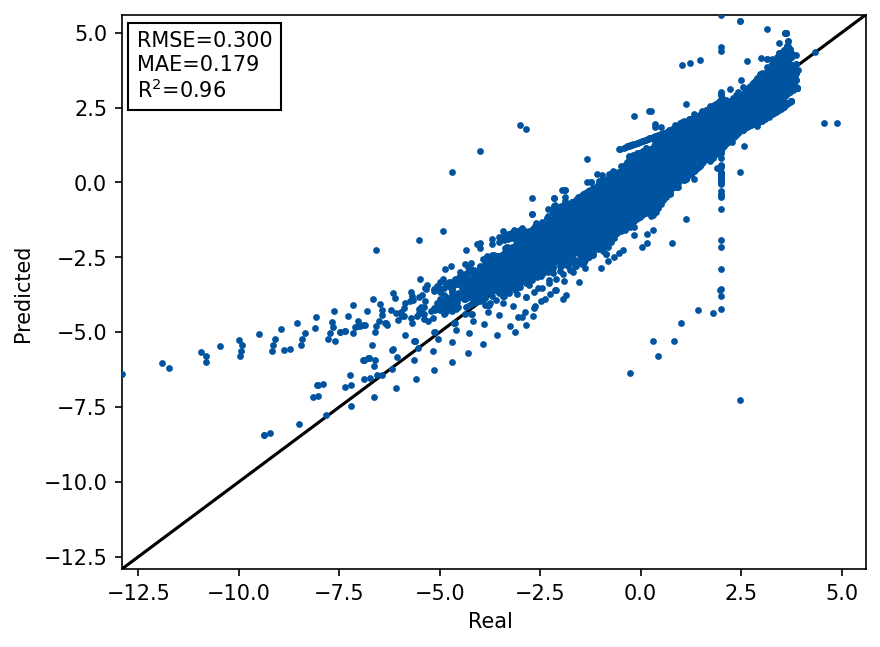}
		\caption{$\text{GNN}_{\text{concat}}$}
	\end{subfigure}
	\begin{subfigure}[t]{0.33\textwidth}
		\centering
		\includegraphics[width=\linewidth]{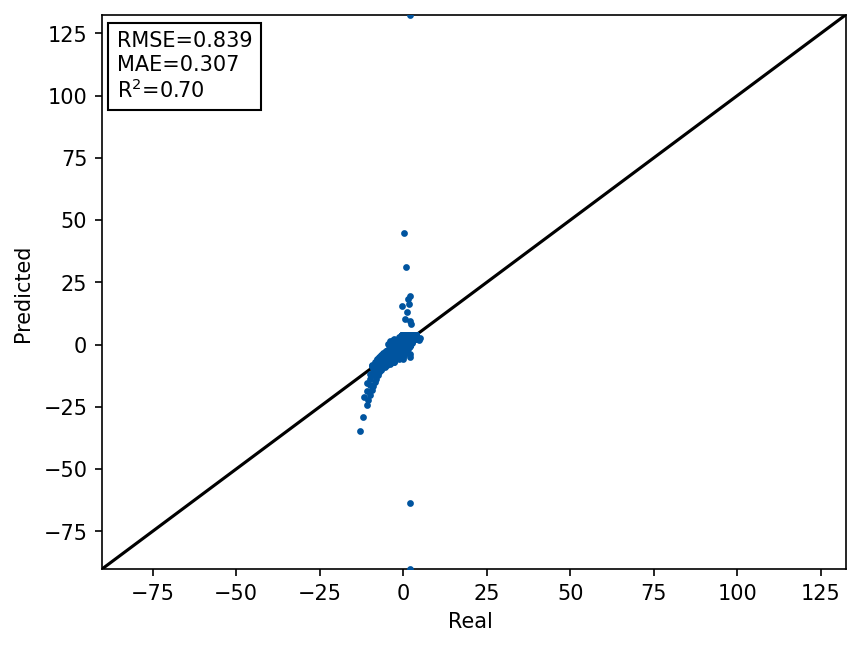}
		\caption{$\text{GNN}_{\text{Antoine}}$}
		\label{fig:par_plot_singleT_GNN_ant}
	\end{subfigure}
	
	\caption{Parity plots of test set $\text{p}^{sat}_{SingleT}\text{(T)}$}
	\label{fig:parity_p_sTpm}
\end{figure}

\noindent Next, in Figure~\ref{fig:parity_p_sTpm}, we consider the parity plots for the $\text{p}^{sat}_{SingleT}\text{(T)}$ case, where only a single, randomly selected temperature point per molecule is used for training (cf. Section~\ref{sec:case_studies}); again showing the results for the best out of ten runs.

We see that the DeepEOSNet predictions are closely aligned along the diagonal, with a slight tendency to overestimate in the low vapor pressure region.
The GNN$_\text{concat}$ overestimates the vapor pressures in both the low and high pressure region.
This reflects the difference in average RMSE between the DeepEOSNet with 0.31 and the GNN$_\text{concat}$ with 0.34 (see Table~\ref{tab:perform_met_vap_pres}).
In general, both have a more widened distribution along the diagonal than in the $\text{p}^{sat}_{extra}\text{(T)}$ scenario, which can be explained by the fact that much less data is used for training, in fact, only a single temperature data point per molecule compared to 80\% of available data.
The best GNN$_\text{Antoine}$ performs significantly worse in the $\text{p}^{sat}_{SingleT}\text{(T)}$ scenario, with a strong tendency to underestimate in the lower vapor pressure regions and overall numerous structural outliers. 
This is also reflected in the high average RMSE, which is an order of magnitude larger compared to the $\text{p}^{sat}_{extra}\text{(T)}$ scenario.
We thus hypothesize that the limited amount of data available for training in the $\text{p}^{sat}_{SingleT}\text{(T)}$ scenario does not suffice to learn suitable Antoine parameters.
To a certain extent this is expected as in traditional parameter estimation the Antoine parameters would not be identifiable with a single temperature point per molecule.
Thus, we focus on the DeepEOSNet and GNN$_\text{concat}$ in the following.

\begin{figure}[htpb]
	\centering
	
	\begin{subfigure}[t]{0.48\textwidth}
		\centering
		\includegraphics[width=\linewidth]{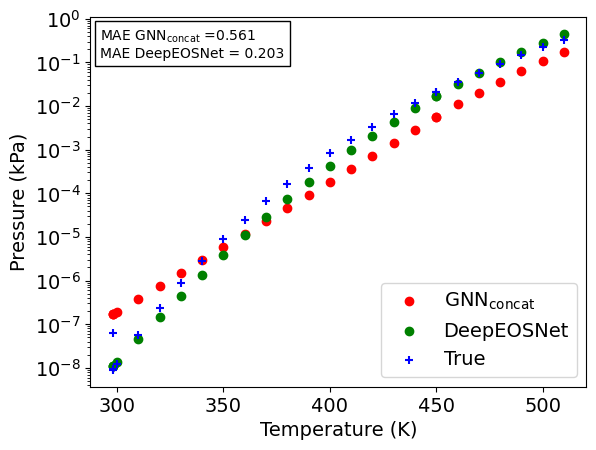}
		\caption{Benzo(k)fluoranthene}
		
	\end{subfigure}
	\hfill
	\begin{subfigure}[t]{0.48\textwidth}
		\centering
		\includegraphics[width=\linewidth]{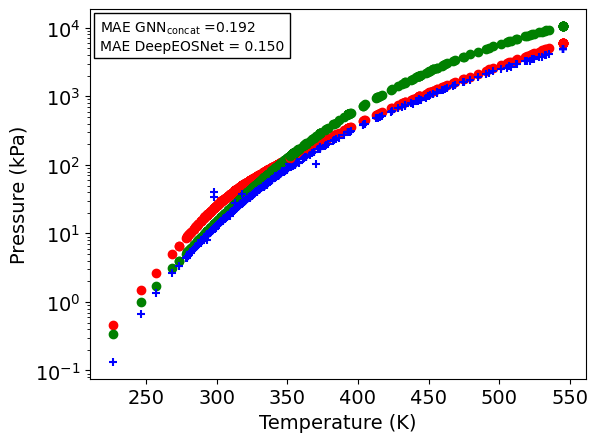}
		\caption{Acetonitrile}
	\end{subfigure}
	
	\begin{subfigure}[t]{0.48\textwidth}
		\centering
		\includegraphics[width=\linewidth]{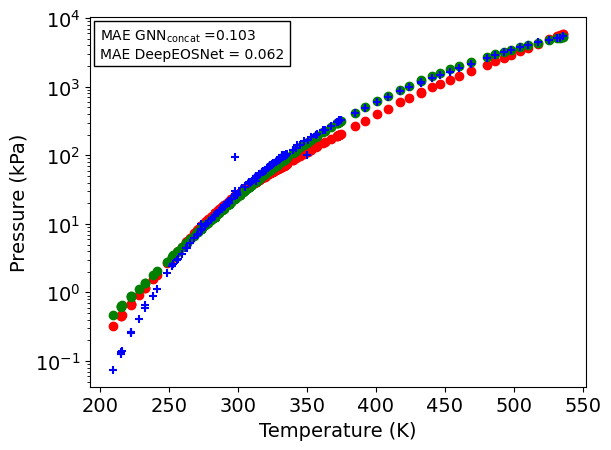}
		\caption{Chloroform}
		\label{fig:vap_pres_chloro}
		
	\end{subfigure}
	\hfill
	\begin{subfigure}[t]{0.48\textwidth}
		\centering
		\includegraphics[width=\linewidth]{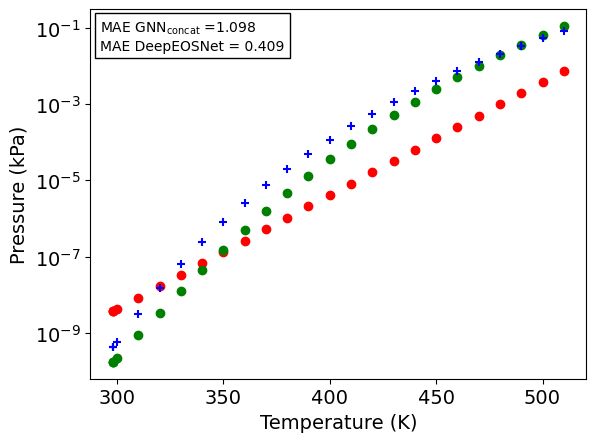}
		\caption{Benzo(g,h,i)perylene}
	\end{subfigure}
	
	\begin{subfigure}[t]{0.48\textwidth}
		\centering
		\includegraphics[width=\linewidth]{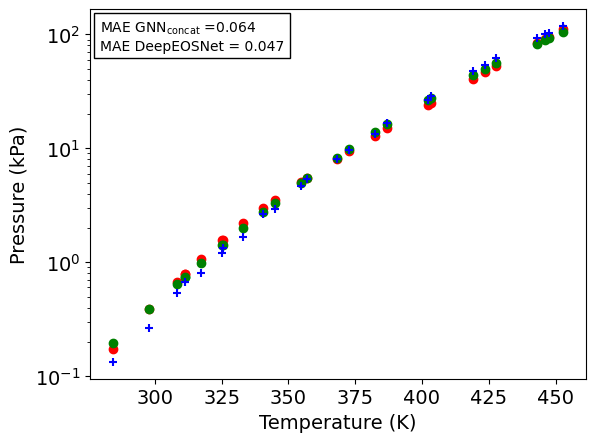}
		\caption{'1,2-Dichloro-4-(trifluoromethyl)benzene}
		
	\end{subfigure}
	\hfill
	\begin{subfigure}[t]{0.48\textwidth}
		\centering
		\includegraphics[width=\linewidth]{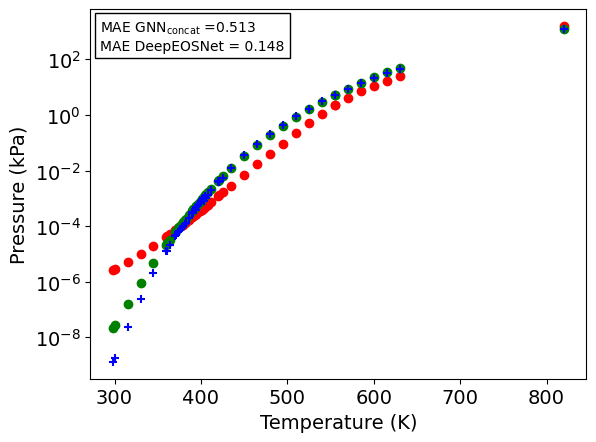}
		\caption{Eicosanoic acid}
	\end{subfigure}
	
	\caption{Vapor pressure curve for sample molecules $\text{p}^{sat}_{SingleT}\text{(T)}$}
	\label{fig:vap_pres_curve_sampleMol}
\end{figure}

In Figure~\ref{fig:vap_pres_curve_sampleMol}, we depict the $p^{sat}(T)$ relation for six sample molecules for the $\text{p}^{sat}_{SingleT}\text{(T)}$ case, which are randomly sampled from the dataset, together with the predictions by the DeepEOSNet and GNN$_\text{concat}$.
For (a) benzo\-(k)fluoranthene and (d) benzo\-(g,h,i)perylene, the GNN$_\text{concat}$ predicts near-linear behavior in the logarithmic vapor pressure scale across the entire temperature domain; for (f) eicosanoic acid, the prediction is nearly linear.
As such, the GNN$_\text{concat}$, fails to capture the nonlinearity in the temperature dependency for these molecules, resulting in systematic over- and underestimations.
The DeepEOSNet model correctly identifies nonlinearities across the whole vapor pressure domain, without any systematic deviations. 
In the other three molecule examples, namely (b) acetonitrile, (c) chloroform and (e) '1,2-dichloro-4-(trifluoromethyl)benzene both models capture the curvature of the experimental data well. 
For (b) acetonitrile, the DeepEOSNet exhibits deviations from the experimental values in the high temperature region, whereas the GNN$_\text{concat}$ deviates in the low temperature region.
For (c) chloroform, both models deviate in the low temperature region, with the GNN$_\text{concat}$ also underestimating slightly at moderate temperatures.
The widened distribution along the diagonal in the parity plot (see Figure~\ref{fig:parity_p_sTpm}) is likely to be attributed to such deviations.
Given that the $\text{p}^{sat}_{SingleT}\text{(T)}$ case has a roughly 99\% smaller training set size than the $\text{p}^{sat}_{extra}\text{(T)}$ case, the prediction accuracy achieved by both models -- in particular of the DeepEOSNet -- is overall highly promising. 

Notably, the individual vapor pressure curves (see Figure~\ref{fig:vap_pres_curve_sampleMol}) also highlight problems with experimental data quality.
Particularly for chloroform (see Figure~\ref{fig:vap_pres_chloro}), the outlier at 300\,K is potentially a conversion error, as an order of magnitude smaller would better align with the remainder of the dataset.
This highlights that even curated experimental databases, e.g., as here provided by NIST, can include erroneous data points, which might explain at least some of the outliers observed in the parity plots (see Figure~\ref{fig:parity_p_text}).
Such erroneous data points occur frequently in molecular property datasets, are difficult to verify and therefore represent a practical scenario.
Additionally, we are rather interested in comparing different methods, including how they are affected in such cases, hence we decided to keep the dataset as provided by NIST.

\subsection{Mixture molar volume prediction}
\label{sec:mol_v_res}

\noindent Lastly, we consider the $\text{V}^{mix}_{SingleX}\text{(T,p,x)}$ case, i.e., prediction of the molar volume of binary mixtures as a function temperature, pressure, and composition (cf. Section~\ref{sec:case_studies}).
Here, we compare the DeepEOSNet to the GNN$_\text{concat}$ architecture, i.e., to simply concatenating the three state variables to the binary mixture fingerprint vector.

\begin{table}[htpb]
	\centering
	\caption{Performance metrics, root mean squared error (RMSE), mean absolute error (MAE), and coefficient of determination ($\text{R}^{2}$) evaluated on the logarithmic scale, of different prediction models on test set for the mixture molar volume prediction scenario.}
	\label{tab:perform_met_v_mol}
	\begin{tabular}{c c|c c c|c c c}
		
		\multirow{2}{*}{Scenario} & & \multicolumn{3}{c|}{$\text{GNN}_{\text{concat}}$} & \multicolumn{3}{c}{DeepEOSNet} \\
		& & RMSE & MAE & $\text{R}^{2}$ & RMSE & MAE & $\text{R}^{2}$\\
		\hline
		
		\multirow{2}{*}{$\text{V}^{mix}_{SingleX}\text{(T,p,x)}$}& avg & 0.013 & 0.0050 & 0.99 & 0.015 & 0.0064 & 0.99 \\
		& $\sigma$ & 0.0048 & 0.0036 & 0.0073 & 0.0076 & 0.0058 & 0.015 \\
		\hline
	\end{tabular}
\end{table}

In Table~\ref{tab:perform_met_v_mol}, we show the prediction metrics for both models.
Both the DeepEOSNet and the GNN$_\text{concat}$ achieve good overall accuracies with average $\text{R}^{2}$ values at 0.99, and low RMSEs and MAEs in a similar range.
The relatively low standard deviation obtained from ten different training runs indicates stable training, as in the vapor pressure case studies; notably, we also find similar convergence rates for both models.

\begin{figure}[h]
	\centering
	\begin{subfigure}[t]{0.45\textwidth}
		\centering
		\includegraphics[width=\linewidth]{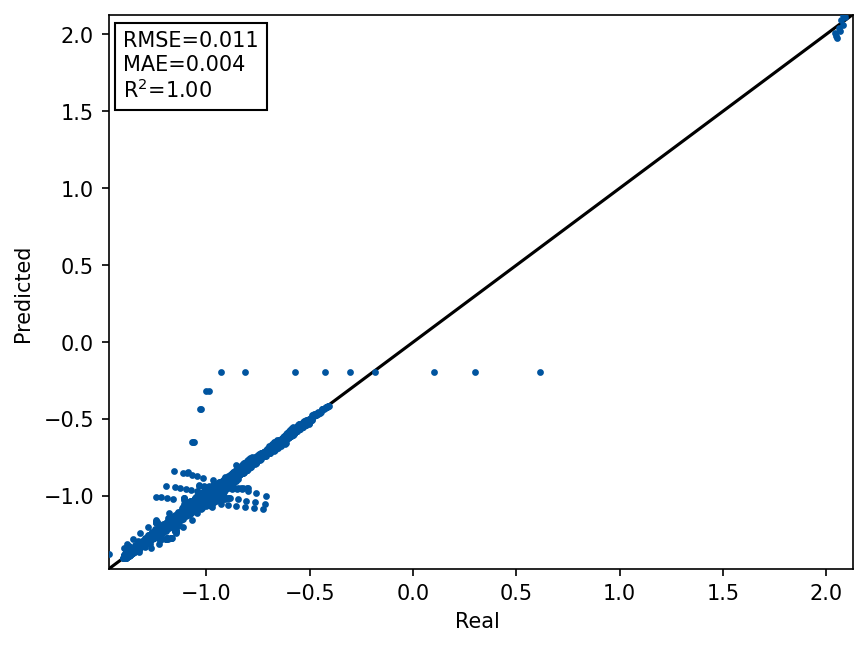}
		\caption{DeepEOSNet}
	\end{subfigure}
	\hfill
	\begin{subfigure}[t]{0.45\textwidth}
		\centering
		\includegraphics[width=\linewidth]{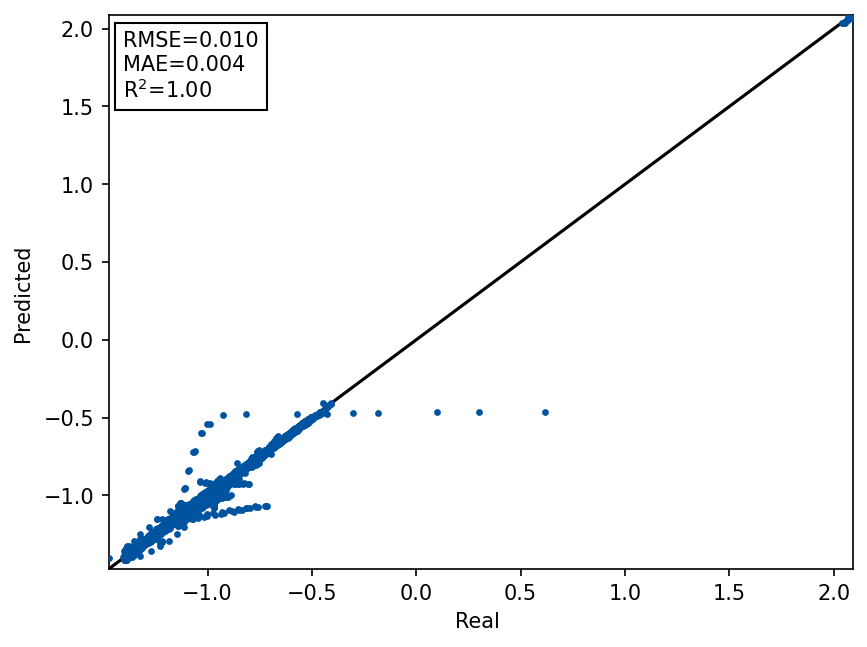}
		\caption{$\text{GNN}_{\text{concat}}$}
	\end{subfigure}
	\caption{Parity plots of test set $\text{V}^{mix}_{SingleX}\text{(T,p,x)}$}
	\label{fig:par_plot_vmix}
\end{figure}

We further show the parity plots for the $\text{V}^{mix}_{SingleX}\text{(T,p,x)}$ scenario in Figure~\ref{fig:par_plot_vmix}.
For both models, a large fraction of the predictions matches the diagonal closely, emphasizing the high prediction accuracy.
Given that the $\text{V}^{mix}_{SingleX}\text{(T,p,x)}$ is a more high-dimensional prediction task, which is inherently more challenging, both models show good overall prediction performance on a similar level.
However, for both models, systematic misalignment for some groups of points is visible, almost in a horizontal alignment. 
Such prediction behavior can indicate that one or multiple state dependencies are not learned correctly but systematically over- or underestimated, which we investigate by means of individual molecules in the following.

\begin{figure}[h]
	\centering
	
	\begin{subfigure}[t]{0.48\textwidth}
		\centering
		\includegraphics[width=\linewidth]{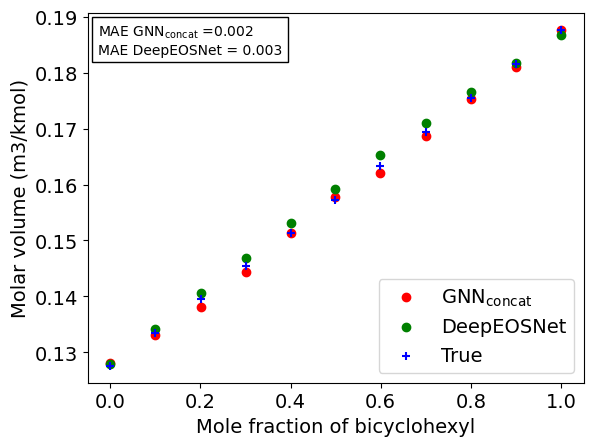}
		\caption{Bicyclohexyl \& Methylcyclohexane at 293.15\,K and 100.8\,kPa}
		\label{fig:mol_v_ind_a}
		
	\end{subfigure}
	\hfill
	\begin{subfigure}[t]{0.48\textwidth}
		\centering
		\includegraphics[width=\linewidth]{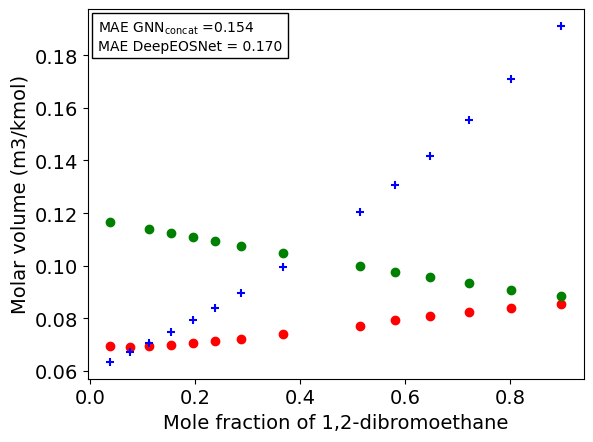}
		\caption{1,2-Dibromoethane \& Butyl propanoate at 298.15\,K and 101\,kPa}
		\label{fig:mol_v_ind_b}
	\end{subfigure}
	
	\begin{subfigure}[t]{0.48\textwidth}
		\centering
		\includegraphics[width=\linewidth]{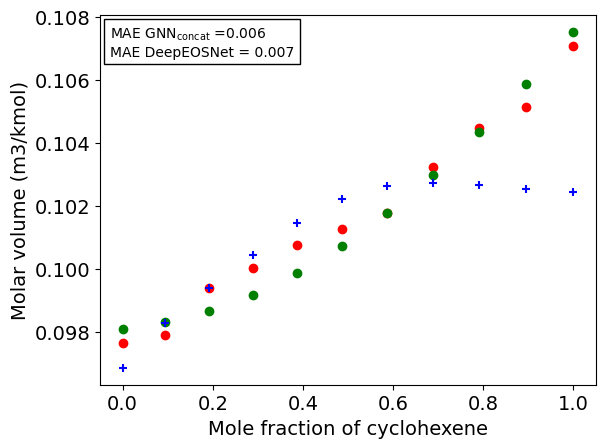}
		\caption{Cyclohexene \& N-methyl-2-pyrrolidone at 303.15\,K and 101\,kPa}
		\label{fig:mol_v_curve_azeotrope}
		
	\end{subfigure}
	\hfill
	\begin{subfigure}[t]{0.48\textwidth}
		\centering
		\includegraphics[width=\linewidth]{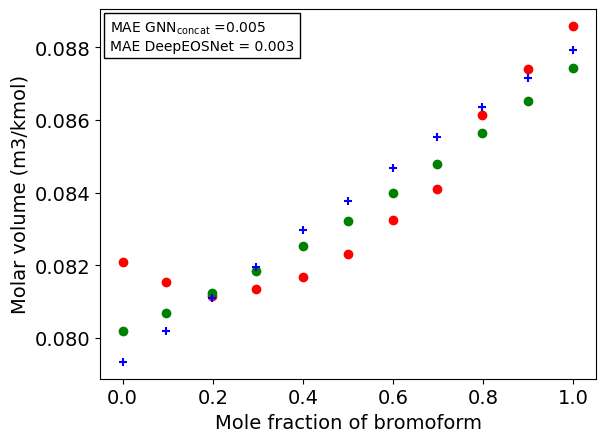}
		\caption{Bromoform \& 2-Methoxyethanol at 298.15\,K and 101.325\,kPa}
		\label{fig:mol_v_ind_d}
	\end{subfigure}
	
	\caption{Molar volume over mole fractions for four exemplary mixtures in the $\text{V}^{mix}_{SingleX}\text{(T,p,x)}$ case}
	\label{fig:mol_v_curve}
\end{figure}

Figure~\ref{fig:mol_v_curve} shows the molar volume over the composition of four individual binary mixtures, where (a) is selected randomly and the other three are selected to show different types of prediction difficulties. 
We choose the composition as independent variable, as we only include one composition point per binary mixture in the training (but at varying temperature and pressure) in the $\text{V}^{mix}_{SingleX}\text{(T,p,x)}$ case, hence learning the composition dependency is presumably most challenging.

For the mixture of (a) bicyclohexyl \& methylcyclohexane, which behaves close to an ideal mixture with near-linear composition dependency, both the GNN$_\text{concat}$ and the DeepEOSNet capture the trends well. 
For the mixture of (b) 1,2-dibromoethane \& butyl propanoate both models fail to capture the composition dependency correctly, despite a rather simple near-linear relation. 
This is also reflected in the horizontal lines in the parity plots (see Figure~\ref{fig:par_plot_vmix}) as described above. 
This highlights that even for some mixtures with near-ideal behavior, a single data point in the composition domain is not sufficient for capturing the composition dependency accurately.

For the mixture of (c) cyclohexene \& n-methyl-2-pyrrolidone, which form an azeotrope, both models are not capable to approximate the composition dependency and its local maximum accurately.
Hence for this mixture, a single data point in the composition domain is not sufficient for accurately predicting azeotropes for neither of the two models.
We note that less than 2\% of mixtures in the dataset are azeotropes, which explains the tendency of both models to predict zeotropic mixtures.
Contrary, for the mixture of (d) bromoform \& 2-methoxyethanol example, the GNN$_\text{concat}$ falsely predicts an azeotrope, whereas the DeepEOSNet more accurately captures the near-linear curvature.
Examples (b), (c) and (d) explain the misalignment for specific points that we observe in the parity plots (Figure~\ref{fig:par_plot_vmix}).
Specifically, the last two mixture examples stress that azeotropes pose a significant challenge to both models in a data scarce scenario, and it remains unclear which model is more suitable for predicting them. 
As the accurate prediction of azeotropes has high implications on chemical engineering applications, these limitations should be considered when using the models in practice. 

In general, the mixture examples in Figure~\ref{fig:mol_v_curve} highlight the added difficulty that the $\text{V}^{mix}_{SingleX}\text{(T,p,x)}$ scenario entails: The composition dependency of the molar volume differs for the individual mixtures across the dataset.
This is structurally different to the first two case studies focusing on the vapor pressure, as the temperature dependency of the vapor pressure typically has a similar form for all molecules~\cite{pfennig2013thermodynamik}. 
The advantage of the DeepEOSNet observed in the $\text{p}^{sat}_{SingleT}\text{(T)}$ case seems to appear in scenarios where data is sparse in the state domain \emph{and} the state dependency is structurally similar across the entire dataset.
Overall, we therefore find the DeepEOSNet architecture to result in at least similar prediction accuracy compared to current approaches in capturing state dependencies, additionally providing benefits for properties for which the state dependency is similar between molecules. 
DeepEOSNet is thus a promising alternative for predicting the state dependencies of molecular and mixture properties. 

\section{Conclusion}
\label{sec:conc}

\noindent We propose an architectural adaptation to GNNs for predicting the state-dependent thermodynamic properties of molecules. 
The proposed architecture, DeepEOSNet, takes inspiration from DeepONets and SplitNets, i.e., the prediction is separated into two network channels, capturing molecular structure and general state dependencies, respectively.
We apply DeepEOSNet to three property prediction case studies and compare its performance to simply concatenating state information to molecular fingerprints and embedding semi-empirical thermodynamic equations. 
In the first case study, where vapor pressure is to be predicted and data is abundant along the temperature domain, both the DeepEOSNet and GNN$_\text{concat}$ perform equally well, outperforming the GNN with embedded semi-empirical equations in terms of training stability. 
In case of data sparsity along the temperature domain of vapor pressures, which is tested in the second case study, where only a single temperature point per molecule is used for training, the DeepEOSNet shows the highest prediction performance.
This highlights the capabilities of the DeepEOSNet to learn functional relations, here state dependencies, in a data efficient way. 
The results from the third case study, that is, predicting the molar volume of binary mixtures as a function of temperature, pressure, and composition, show that the DeepEOSNet's capabilities can be limited to prediction tasks where the dependency on the additional input feature is structurally similar across the dataset.

In summary, DeepEOSNet is therefore a promising architecture for capturing the state dependencies of thermodynamic properties when training data is sparse in the state domain and the state dependency of the property of interest is structurally similar across the dataset.

Future work can easily transfer DeepEOSNet to predicting further state-dependent molecular and mixture properties.
Here, a promising extension is the combination of DeepEOSNet with thermodynamics-consistent architectures \cite{rittig2024thermodynamics_consistent}.

\section*{CRediT authorship contribution statement}
\textbf{Jan Pav\v{s}ek:} Conceptualization, Methodology, Software, Formal analysis, Investigation, Writing - original draft, Visualization.
\textbf{Alexander Mitsos:} Conceptualization, Writing - review \& editing, Supervision,  Funding Acquisition.
\textbf{Manuel Dahmen:} Conceptualization, Writing - review \& editing.
\textbf{Tai Xuan Tan:} Methodology, Writing - review \& editing.
\textbf{Jan G. Rittig:}
Conceptualization, Software, Writing - original draft, Supervision, Funding Acquisition.

\section*{Data and Software Availability}

\noindent The data used for training of the presented models are confidential and can be accessed through the NIST Thermodata engine. The code is available in our \emph{GitLab} repository \href{https://git.rwth-aachen.de/avt-svt/public/gmolprop/}{\emph{GMoLprop}}.

\section*{Acknowledgements}
\noindent This project was funded by the Deutsche Forschungsgemeinschaft (DFG, German Research Foundation) – 466417970 – within the Priority
Programme ‘‘SPP 2331: Machine Learning in Chemical Engineering’’. 

Alexander Mitsos and Manuel Dahmen received funding from the Helmholtz Association of German Research Centers. 

Model training was performed with computing resources granted by RWTH Aachen University.

The authors thank René Görgen for his software engineering support.

\pagebreak

\bibliographystyle{unsrt}  
\bibliography{references}  

\end{document}